\documentclass{osa-article}

\journal{osajournal}

\setprjcopyright

\articletype{Letter}

\begin{document}

\title{Mid-infrared photon counting and resolving via efficient frequency upconversion}

\author{Kun Huang,\authormark{1,*} Yinqi Wang,\authormark{1}  Jianan Fang,\authormark{1}  Weiyan Kang,\authormark{2} Ying Sun,\authormark{2} Yan Liang,\authormark{2} Qiang Hao,\authormark{2} Ming Yan,\authormark{1} and Heping Zeng\authormark{1,3,4,$\dagger$}}

\address{\authormark{1}State Key Laboratory of Precision Spectroscopy, East China Normal University, Shanghai 200062, China\\
\authormark{2}School of Optical Electrical and Computer Engineering, University of Shanghai for Science and Technology, Shanghai 200093, China\\
\authormark{3}Jinan Institute of Quantum Technology, Jinan, Shandong 250101, China\\
\authormark{4}Shanghai Research Center for Quantum Sciences, Shanghai 201315, China}

\email{\authormark{*}khuang@lps.ecnu.edu.cn\\
\authormark{$\dagger$}hpzeng@phy.ecnu.edu.cn} 


\begin{abstract}
Optical detectors with single-photon sensitivity and large dynamic range would facilitate a variety of applications. Especially, the capability of extending operation wavelengths into the mid-infrared region is highly attractive. Here we implement a mid-infrared frequency upconversion detector for counting and resolving photons at 3 $\mu$m. Thanks to the spectro-temporal engineering of the involved optical fields, the mid-infrared photons could be spectrally translated into the visible band with a conversion efficiency of 80\%. In combination with a silicon avalanche photodiode, we obtained unprecedented performances with a high overall detection efficiency of 37\% and a low noise equivalent power of 1.8$\times$10$^{-17}$ W/Hz$^{1/2}$. Furthermore, photon-number-resolving detection at mid-infrared wavelengths was demonstrated, for the first time to our knowledge, with a multi-pixel photon counter. The implemented upconversion detector exhibited a maximal resolving photon number up to 9 with a noise probability per pulse of 0.14\% at the peak detection efficiency. The achieved photon counting and resolving performance might open up new possibilities in trace molecule spectroscopy, sensitive biochemical sensing, and free-space communications, among others. 
\end{abstract}

\section{Introduction}
The mid-infrared (MIR) wavelength region is of great interest for a variety of applications as wide-ranging as environmental monitoring, molecular spectroscopy, biomedical sensing and free-space communication \cite{EbrahimZadeh2008Book, Vodopyanov2020Book}. In these envisioned scenarios, sensitive MIR detection is highly demanded to access dramatically improved performances in terms of detection sensitivity, working distance or imaging functionality \cite{Razeghi2014RPP}. So far, tremendous progress has been witnessed in developing MIR detectors based on either conventional semiconductors of indium antimonide (InSb) and mercury cadmium telluride (MCT), or emerging materials of colloidal quantum dots \cite{Keuleyan2011NP}, graphene plasmons \cite{Guo2018NM} and black phosphorus \cite{Bullock2018NP}. However, the noise-equivalent power (NEP) is typically limited to about pW/Hz$^{1/2}$, thus far below the level of single-photon detection. The attainable sensitivity could be improved with cryogenic operation for suppressing the intrinsic black-body radiation and dark current. Notably, MIR single-photon detection has be accessed by superconducting nanowire detectors \cite{Marsili2012NL}, albeit with reduced detection efficiency about 4\% and additional complexity of bulky cooling system. Therefore, continuous endeavors have still been dedicated to approaching efficient direct detection for MIR photons.

Indeed, the performance of currently existing MIR detectors is in marked contrast to the visible and near-infrared regions, in which highly-efficient and low-noise photon counting is available by commercial avalanche photodiodes (APDs) \cite{Hadfield2009NP}. Especially, photon-number-resolving (PNR) capability has also been readily achieved using silicon-based multi-pixel photon counter (Si-MPPC). The resulting single-photon sensitivity and large dynamic range could not only support accurate and rapid characterization of non-repetitive optical signal at low-light levels, but also facilitate quantum-optics experiments to investigate multi-photon quantum states \cite{Gerrits2010PRA, Namekata2010NP}, measure high-order correlation functions \cite{Avenhaus2010PRL} and characterize Wigner functions \cite{Nehra2019Optica}. Moreover, photon-number identification also constitutes a key enabler in promising protocols in quantum information science, such as thresholded laser ranging \cite{Cohen2019PRL} and loss-tolerant optical communication \cite{Becerra2015NP}. 

In view of the attractive features of Si-based detectors, frequency upconversion strategy by spectrally translating the MIR signals into shorter wavelengths has been recognized as a simple yet effective way to capture MIR photons. Recently, such indirect MIR approach has been investigated by utilizing optical nonlinearities based on crystals \cite{Temporao2006OL, Zhou2013APL, Mancinelli2017NC, Mrejen2020LPR}, waveguides \cite{Neely2012OL}, nanowires \cite{Liu2012NP}, as well as fluorophore \cite{Zheng2013NP}. Similar spirit was also manifested in MIR detection based on non-degenerate two-photon absorption in wide-bandgap semiconductors \cite{Fishman2011NP}. Current advances in frequency upconversion detectors have even led to demonstrating sensitive MIR imaging by high-definition Si-based CCD cameras \cite{Dam2012NP ,Knez2020LSA}, meanwhile with great potential to extend the working wavelength into the far-infrared regime \cite{Demur2017OL}. However, to date, the conversion efficiency achieved for MIR single-photon upconverters are typically below 35\% \cite{Zhou2013APL, Dam2012NP, Mrejen2020LPR}. These limited values contrast to the near-infrared counterparts with nearly complete conversion \cite{Pelc2011OE, Huang2012APL, Xiang2018PRA}. Moreover, stringent filtering stages are typically needed to suppress the severe pump-induced parametric fluorescence noise, which ultimately result in a MIR detection efficiency of several percentages \cite{Pedersen2019PTL}. Consequently, PNR detection at MIR has not yet been reported due to the limited efficiency and accompanied background noise, although the desirable feature was demonstrated a decade ago at 1.56 $\mu$m \cite{Pomarico2010OE} and 1.04 $\mu$m \cite{Huang2011OL}. Nowadays, there is a significant impulse to move quantum optics to the MIR \cite{McCracken2018JOSAB, Mancinelli2017NC, Sua2017SR}, thus urgently calling for enabling techniques to yield efficient single-photon counting and photon-number resolution.

Here we demonstrate a high-performance MIR frequency upconversion detector based on the coincidence-pumping configuration. Thanks to the spectro-temporal optimization for the passively synchronized pulses, the intrinsic conversion efficiency for the MIR photons at 3 $\mu$m reached to 80\%. Correspondingly, the overall detection efficiency about 37\% could be achieved due to the effective filtering system. The resultant noise equivalent power was as low as 1.8$\times$10$^{-17}$ W/Hz$^{1/2}$, thus representing highest sensitivity among reported MIR detectors. Furthermore, we demonstrated for the first time, to the best of our knowledge, the ability to resolve the number of photons within a MIR pulse. The obtained MIR photon counting and resolving performances would pave the way toward implementing advanced protocols, which require an ultimate sensitivity at the single-photon level and a large dynamic range with linear response to incident photon numbers.

\section{Principle of coherent frequency conversion}
Coherent frequency up-converters enable to translate the frequency of signal photons into a higher targeted one, while preserving the optical coherences or quantum characteristics. The involved nonlinear process is usually realized by the second-order sum-frequency generation (SFG) under the non-depleted pump approximation. The energy conservation leads to $\omega_1 + \omega_2 = \omega_3$, where 1, 2, 3 denote the signal, pump and SFG modes, respectively. Under the perfect phase-matching condition, the relevant Hamiltonian for the three-wave mixing is given as
\begin{equation}
\hat{H} = i \hbar g \alpha (\hat{a}_1\hat{a}_3^\dagger - H.c.)  \ ,
\label{eq1}
\end{equation}
where $\alpha$ is the electric amplitude of the classical pump field, $g$ is the coupling constant determined by the nonlinear susceptibility of the medium, and $H.c.$ represents the Hermitian conjugate. $\hat{a}_1$ and $\hat{a}_3 $ denote the annihilation operators for the signal and converted photons, respectively. Using this Hamiltonian, the state evolution after an interaction length $L$ can be described by the coupled-mode equations as
\begin{equation}
\hat{a}_3(L) = \sin(|g \alpha| L) \hat{a}_1(0) +  \cos(|g \alpha| L) \hat{a}_3(0) \ .
\label{eq2}
\end{equation}
A complete conversion $\hat{a}_3(L)  = \hat{a}_1(0) $ can be obtained if $|g \alpha| L = \pi/2$. In this case, the signal photon is annihilated as the creation of the SFG photon. The one-to-one correspondence ensures a noise-less conversion, which forms the basis for many applications such as single-photon detection and optical quantum interface. Generally, the conversion efficiency can be defined by 
\begin{equation}
\eta = N_3(L)/N_1(0) = \eta_m \sin^2(|g \alpha| L) = \eta_m \sin^2[\pi/2 \sqrt{P/P_m}]  \ , 
\label{eq3}
\end{equation}
where $N_i = \langle \hat{a}_i^\dagger \hat{a}_i \rangle$ denotes the average photon number, $P$ is the  pump power, $\eta_m$ is the maximum conversion efficiency under a pump power of $P_m$.

Similarly, the inverse process acting as a coherent down-converter is possible by noticing the Hermitian conjugated term in Eq. (\ref{eq1}). In this scenario, the signal photon with the higher frequency $\omega_3$ could be completely converted to a photon with a lower frequency $\omega_1$ \cite{Takesue2020PRA}. Therefore, identical solution for the quantum state evolution could be expected, and the conversion efficiency shows the same dependence on the pump power as given by  Eq. (\ref{eq3}). Coherent down-converter was used in our experiment to prepare MIR signal from the near-infrared coherent light. Specifically, the linear response of down-converter provide a solution for precisely calibrating the MIR power especially at the single-photon level. Indeed, measuring MIR light at low powers is typically challenging due to limited powermeter sensitivity and ambient thermal disturbance. To this end, well-calibrated attenuators were inserted in the near-infrared light path, which could result in a same attenuation for the MIR output. In the following, we will turn to the experimental realization of the aforementioned coherent converters.

\section{Experimental setup}
\begin{figure}[t!]
\centering
\includegraphics[width=1 \textwidth]{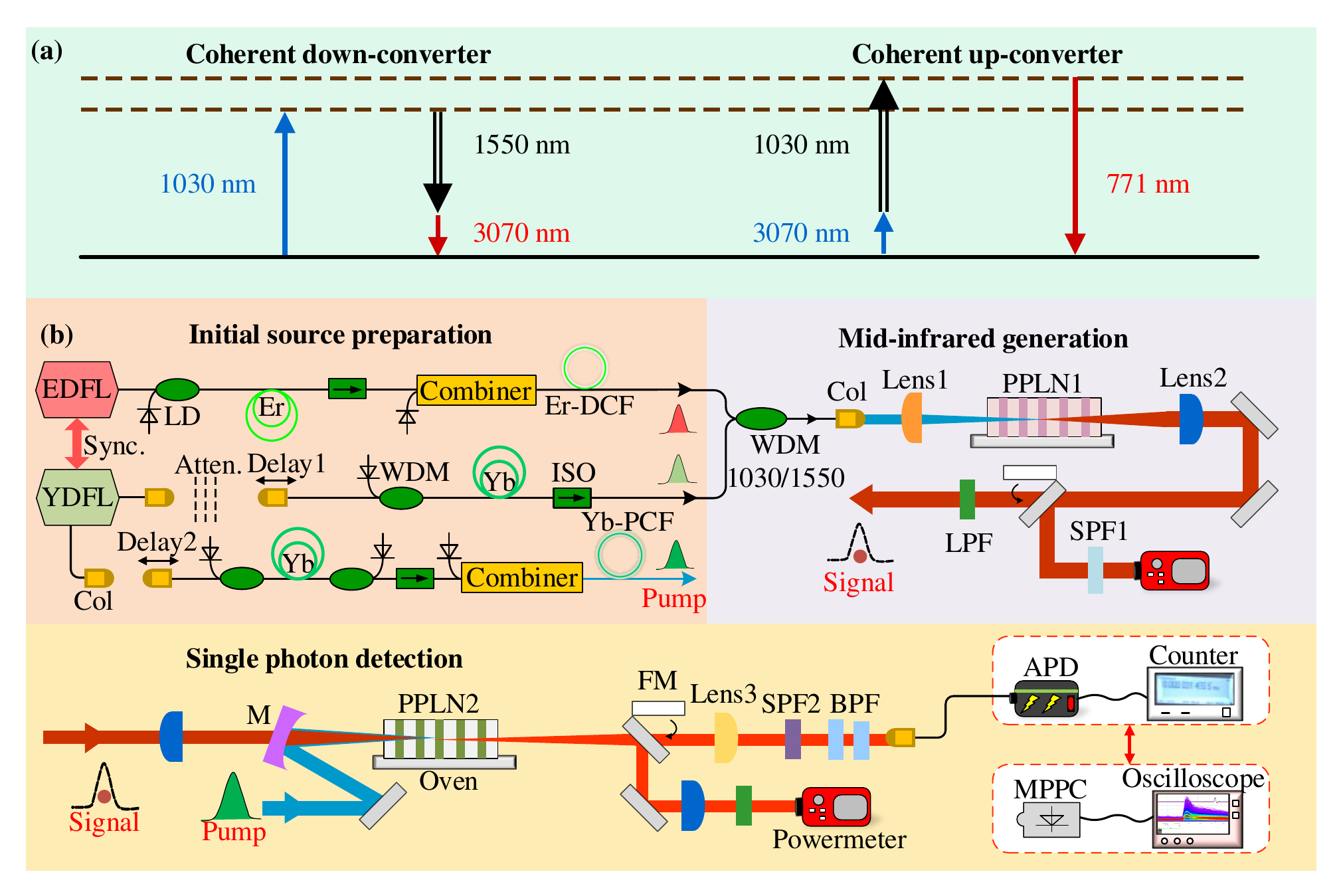}
\caption{(a) Energy transition diagram for coherent down- and up- converters based on second-order nonlinear process. The down-converter is used to prepare a MIR source with a well-calibrated power, which is essential in characterizing the detection performances for the MIR up-converter. (b) Experimental setup for MIR upconversion detection system based on passively synchronized fiber lasers. Initial source preparation for subsequent nonlinear conversion was realized by two Er- and Yb-doped fiber lasers (EDFL and YDFL) and multi-stage fiber amplifiers. The resultant amplified pulses at 1.03 and 1.55 $\mu$m  were used to generate MIR signal. Another branch of amplified pulse at 1.03 $\mu$m served as the pump, which was used to spectrally convert the MIR signal to the visible band for efficient single-photon and PNR detection. LD: laser diode; WDM: wavelength division multiplexer; Col: collimator; ISO: isolator; Yb/Er: ytterbium- or erbium-doped gain fiber; DCF: double-clad fiber; PCF: photonic crystal fiber; Atten: neutral density attenuator; M: dichroic concave mirror; FM: flip mirror; PPLN: periodically-poled lithium niobate crystal; SPF, LPF and BPF: short-, long- and band-pass filter; APD: avalanche photodiode; MPPC: multi-pixel photon counter.}
\label{fig1}
\end{figure}

As discussed above, two complementary nonlinear optical processes, $\textit{i .e.,}$ coherent frequency down- and up- conversion, were used in our experiment. Figure \ref{fig1}(a) illustrates the related energy transition diagram based on three-wave mixing. Specifically, the coherent down-converter (CDC) enables to prepare MIR source from a weak near-infrared light at 1030 nm under the pumping at 1550 nm. In principle, the MIR signal would preserve quantum and coherent properties of the near-infrared states \cite{Huang2013LP}, which thus provides an effective way to prepare quantum states at MIR, such as squeezed states and entangled states. The prepared MIR source here should be in a coherent state, which will be verified latter by the photon-number distribution measured by a PNR detector. Another desirable feature for the CDC was the possibility to calibrate the MIR power particularly at the single-photon level by adding attenuators for the near-infrared light. Indeed, the attenuation for near-infrared wavelengths could be better calibrated due to more sensitive optical detectors. As for the coherent up-converter (CUD), it can spectrally translate the MIR signal into the visible band under pumping at 1030 nm, which enabled to perform MIR single-photon counting and PNR detection.

\begin{figure}[b!]
\centering
\includegraphics[width=0.7\columnwidth]{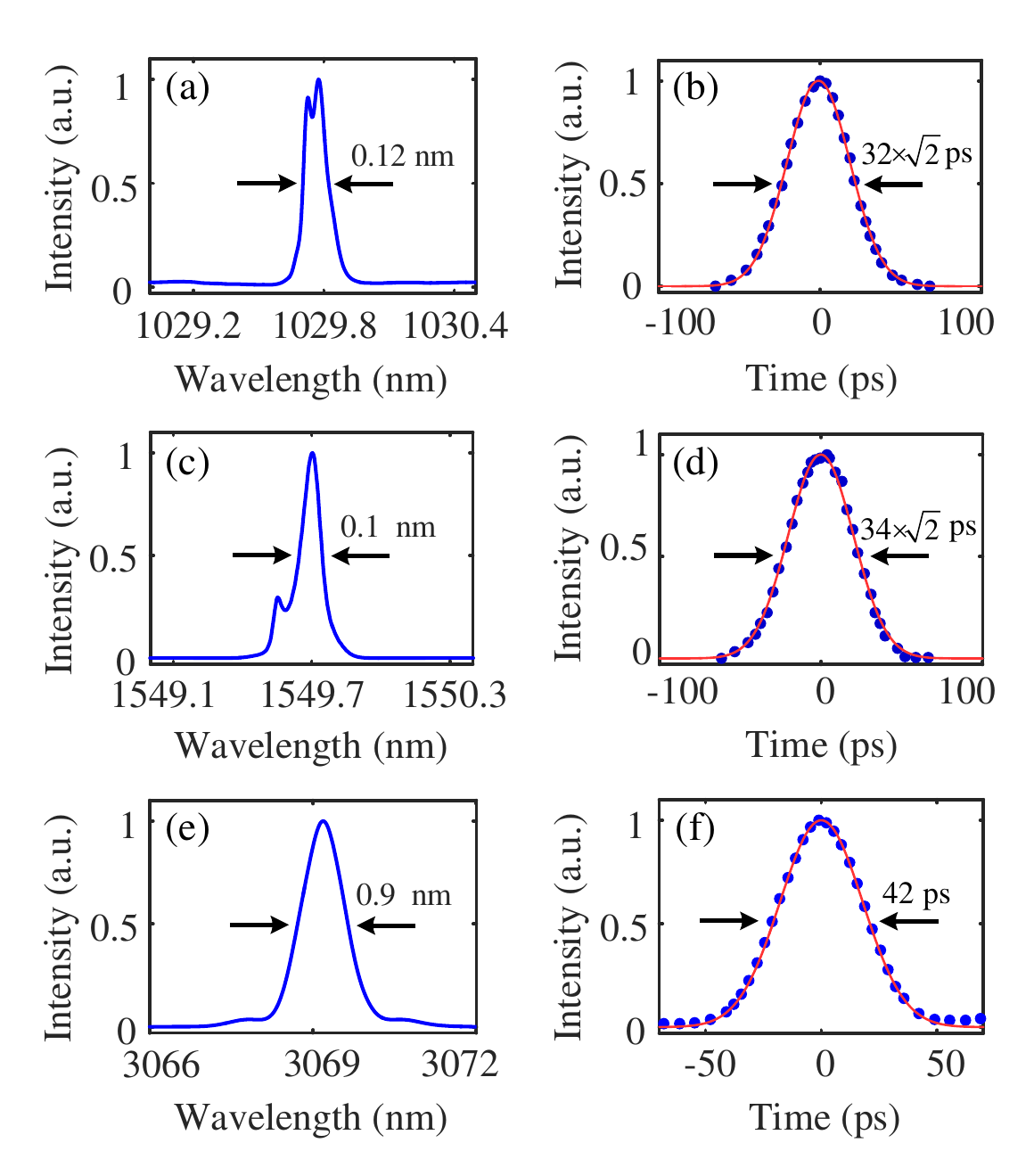}
\caption{Spectro-temporal characterization of involved light pulses. (a) and (c) are the measured optical spectra for the output pulses from the YDFL and EDFL, respectively. (b) and (d) give the corresponding  auto-correlation traces. Scaling factor of $\sqrt{2}$ was taken to deduce the actual pulse duration with the assumption of a Gaussian profile. (e) Optical spectrum for the MIR signal source. (f) Cross-correlation trace between signal and pump pulses.}
\label{fig2}
\end{figure}

The experimental realization was presented in Fig. \ref{fig1}(b), including three parts of initial source preparation, MIR signal generation, and single-photon detection. All the light sources originated from a passively-synchronized fiber laser system, which consisted of two Er- and Yb-doped fiber lasers (EDFL and YDFL). The two fiber lasers were arranged in a shared-cavity configuration. Cross-phase modulation between the dual-color pulses within a common section of single-mode fiber could result in synchronous mode-locking at 14.6 MHz. More details about the all-optical synchronization could be found in \cite{Zeng2019OL}. The YDFL output at 1029.8 nm was divided into two portions, which served as the seed and pump for the CDC and CUC modules, respectively. The power of the pump could be amplified to the watt level based on photonic crystal gain fiber. Thanks to the large mode diameter, a narrow spectrum was maintained during the amplification, resulting in a 0.12-nm full width at half maximum (FWHM) as shown in Fig. \ref{fig2}(a). The corresponding pulse duration was inferred to be 32 ps from the measured auto-correlation trace in Fig. \ref{fig2}(b). Note that a scaling factor of $\sqrt{2}$ was used under the assumption of a Gaussian profile. Similarly, the power of the EDFL output was boosted to about 200 mW by using cascaded fiber amplifiers. The corresponding optical spectrum was shown in Fig. \ref{fig2}(c), which indicated a center wavelength at 1549.7 nm and a FWHM of 0.1 nm. The inferred pulse duration was 34 ps as given in Fig. \ref{fig2}(d).

The dual-color pulses at 1 and 1.55 $\mu$m were then spatially combined with a wavelength division multiplexer (WDM). The mixed light was collimated into the free space before being focused into a periodically-poled lithium niobate (PPLN) crystal. Quasi-phase-matching condition was optimized at an operation temperature of 40.7 $^\circ$C and a polling period of 30.3 $\mu$m, leading to efficient generation of MIR light at 3069 nm. The resultant narrow spectral bandwidth of 0.9 nm as shown in Fig. \ref{fig2}(e) would favor efficient nonlinear conversion within the phase-matching window. To generate MIR signal, the pump power at 1.55 $\mu$m was arbitrary set around 120 mW, corresponding to a down-conversion efficiency of 40\%. The linear relationship between the powers at 1 and 3 $\mu$m was verified by inserting calibrated neutral density filters (NDFs) within the delay line (Delay1). Therefore, the MIR power could be precisely controlled by choosing a proper combination of NDFs at the near infrared. The attenuated MIR light will serve as the signal source for the subsequent CUC module.

To implement the frequency upconversion detector, the signal and pump sources were spatially combined by a dichroic concave mirror and temporally overlapped with a translational stage (Delay2). Another PPLN crystal was used to perform the SFG, which has a length of 25 mm. As shown in the inset of Fig. \ref{fig3}, the phase-matching temperature was about 54.6 $^\circ$C for the chosen  polling period of 20.9 $\mu$m. The resulting SFG signal at 771 nm was then sent into filtering stages before being registered by Si-based detectors. Based on SFG configuration, a cross-correlation trace was measured with a FWHM of 42 ps as shown in Fig. \ref{fig2}(f). In the weak conversion regime, the cross-correlation bandwidth depends on the convolution between the signal and pump pulses, which indicated a pulse duration of 27 ps for the MIR signal. 

\section{Results and discussions}
\begin{figure}[b!]
\centering
\includegraphics[width=0.6\columnwidth]{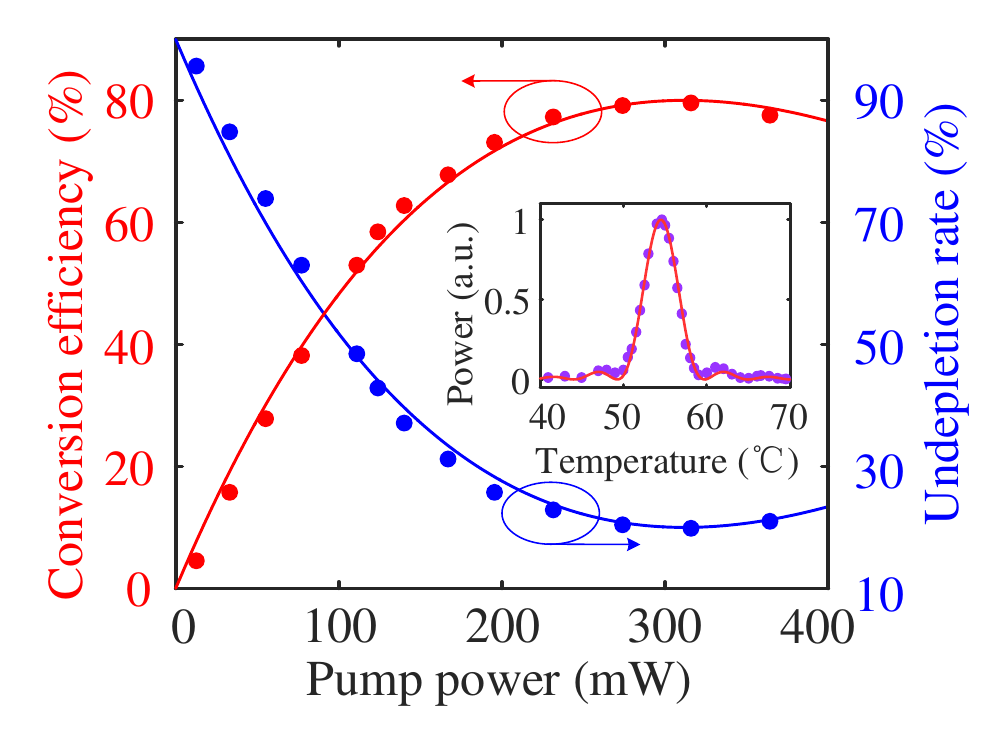}
\caption{Conversion efficiency and undepletion rate as a function of the pump power. The MIR signal was set at mW level, thus ensuring to fulfill the non-depleted pump approximation. The solid lines are given by the theoretical model in the text. Inset shows the phase-matching behavior depending on the crystal temperature.}
\label{fig3}
\end{figure}

Now we turn to characterize the performances of the implemented upconversion detector. Conversion efficiency is a crucial parameter that defines the possibility for a MIR photon to be converted. Practically, the conversion efficiency $\eta$ could be measured by evaluating the power ratio between the signal and upconverted light as $\eta = P_\text{up}/P_s \times \lambda_\text{up}/\lambda_s$. Figure \ref{fig3} shows the deduced conversion efficiency as varying the pump power, which was fitted by Eq. \ref{eq3} with $\eta_m$ = 80\% and $P_m$ = 312 mW. As expected, excessive pumping power would result in the decrease of the conversion efficiency due to the complementary down-conversion process \cite{Pelc2011OE}. To further validate the achieved efficiency, we also measured the so-called undepletion rate $R = 1-\eta = P_\text{res}/P_0$, where $P_0$ and $P_\text{res}$ are the input and residual MIR signal power, respectively. The advantages of the proposed method lie in the avoidance to calibrate power measurement devices for different wavelengths, and the independence from losses including propagation transmission and filtering efficiency. The measured undepletion rate shown in Fig. \ref{fig3} was modeled quite well with the previously obtained parameters. The high conversion efficiency was due to the careful optimization of optical pulses in spectral and temporal domains \cite{Kang2020PTL}. Firstly, the spectrum of MIR signal was made to be narrow for approaching the efficient phase-matching window. Secondly, the pulse duration of the MIR signal was set to be smaller than that of the pump, which ensured sufficiently intensive pump field for signal photons. Thirdly, the relative timing jitter of the synchronized signal and pump pulses was negligible relative to the pulse duration, thus permitting a stable and efficient nonlinear interaction. 

\begin{figure}[t!]
\centering
\includegraphics[width=0.6\columnwidth]{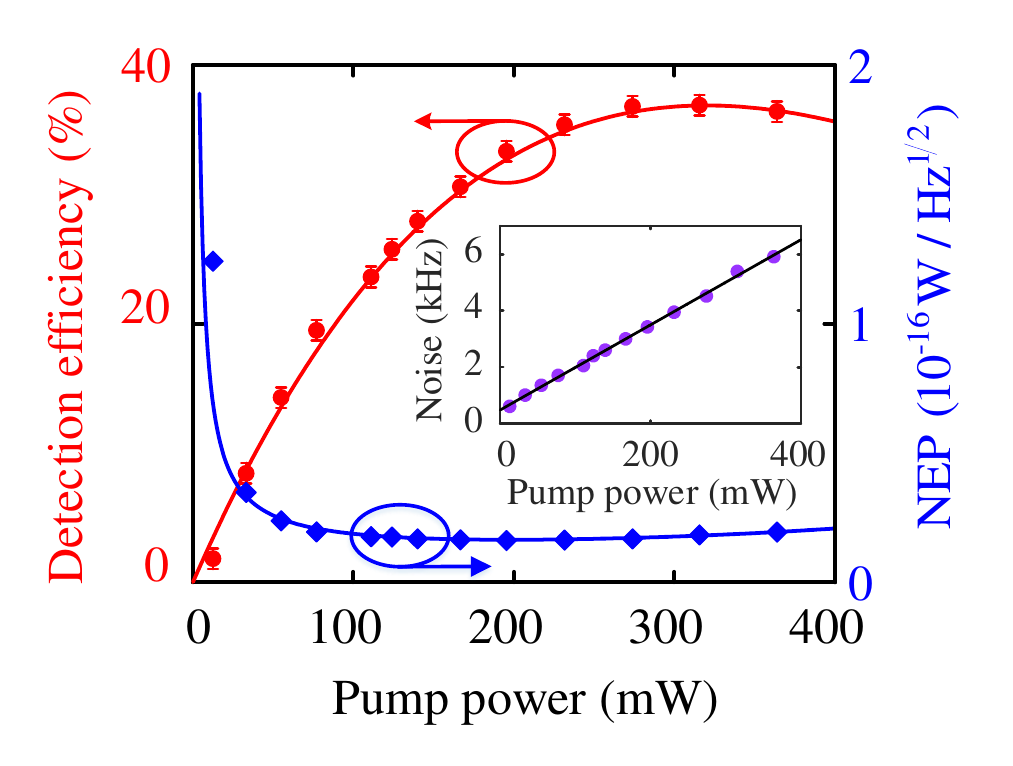}
\caption{Detection efficiency and noise equivalent power (NEP) versus the pump power for the implemented frequency upconversion detector. The input MIR signal was set at the single-photon level with an average photon number of 0.13 per pulse. The solid lines are given by the theoretical model in the text. Count rates for the background noise is given in the inset.}
\label{fig4}
\end{figure}

In the following, we proceed to investigate the photon counting performance for the MIR upconversion detector. To emulate the single-photon source, the MIR signal was intensively attenuated to contain 0.127 photons per pulse. As discussed above, the attenuation was determined by adding a series of properly calibrated neutral density filters into the near-infrared light path. The SFG signal was sent through a short-pass filter with a cut-off wavelength of 900 nm and two band-pass filters with a bandwidth of 10 nm. The total transmission of the filters $\eta_\text{f}$ was about 92\% with a noise rejection of 130 dB at the pump wavelength. The filtered light was then coupled into a single-mode fiber with a coupling efficiency  $\eta_\text{c}$ of 87\%, which could act as an effective spatial filtering for upconverted parametric fluorescences \cite{Meng2018OE}. The resulting background noise $N_b$ at the peak conversion efficiency was about 5.4 kHz as shown in the inset of Fig. \ref{fig4}. Finally, the upconverted photons were registered by a fiber-coupled single-photon counting module (SPCM) with a detection efficiency $\eta_\text{SPCM}$ of 58\% at 771 nm and a dark noise about 100 Hz. The overall detection efficiency could be evaluated by dividing the recored count rate by the input MIR photon flux of 1.854 MHz. Specifically, the maximum count rate of 687 kHz was recorded at the peak conversion efficiency, which indicated a detection efficiency $\eta_d$ of 37\% as shown in Fig. \ref{fig4}. After correction for the total loss given by $\eta_f \times \eta_c  \times \eta_\text{SPCM}$=46\%, the conversion efficiency for MIR single photons was calculated to be 80\%, which was identical to the one obtained in the weak signal regime. Furthermore, sensitivity of the upconversion detector was evaluated with the noise equivalent power defined by NEP=$\hbar \omega \sqrt{2 N_b}/\eta_d$, where $\hbar \omega$ is the MIR photon energy \cite{Pelc2011OE}. As shown in Fig. \ref{fig4}, a low NEP of 1.8$\times 10^{-17} \text{W/Hz}^{1/2}$ was realized at the maximum detection efficiency. The demonstrated sensitivity down to the single-photon level was at least four orders of magnitude better than conventional detectors based on PbS, PbSe, and HgCdTe \cite{Pedersen2019PTL}. Moreover, the NEP for the MIR upconversion detector was more than ten times lower than that for the state-of-the-art superconducting nanowire single-photon detector with a detection efficiency about 4\% at 3 $\mu$m and a dark count rate over 10 kHz \cite{Marsili2012NL}. 

\begin{figure}[t!]
\centering
\includegraphics[width=0.9\textwidth]{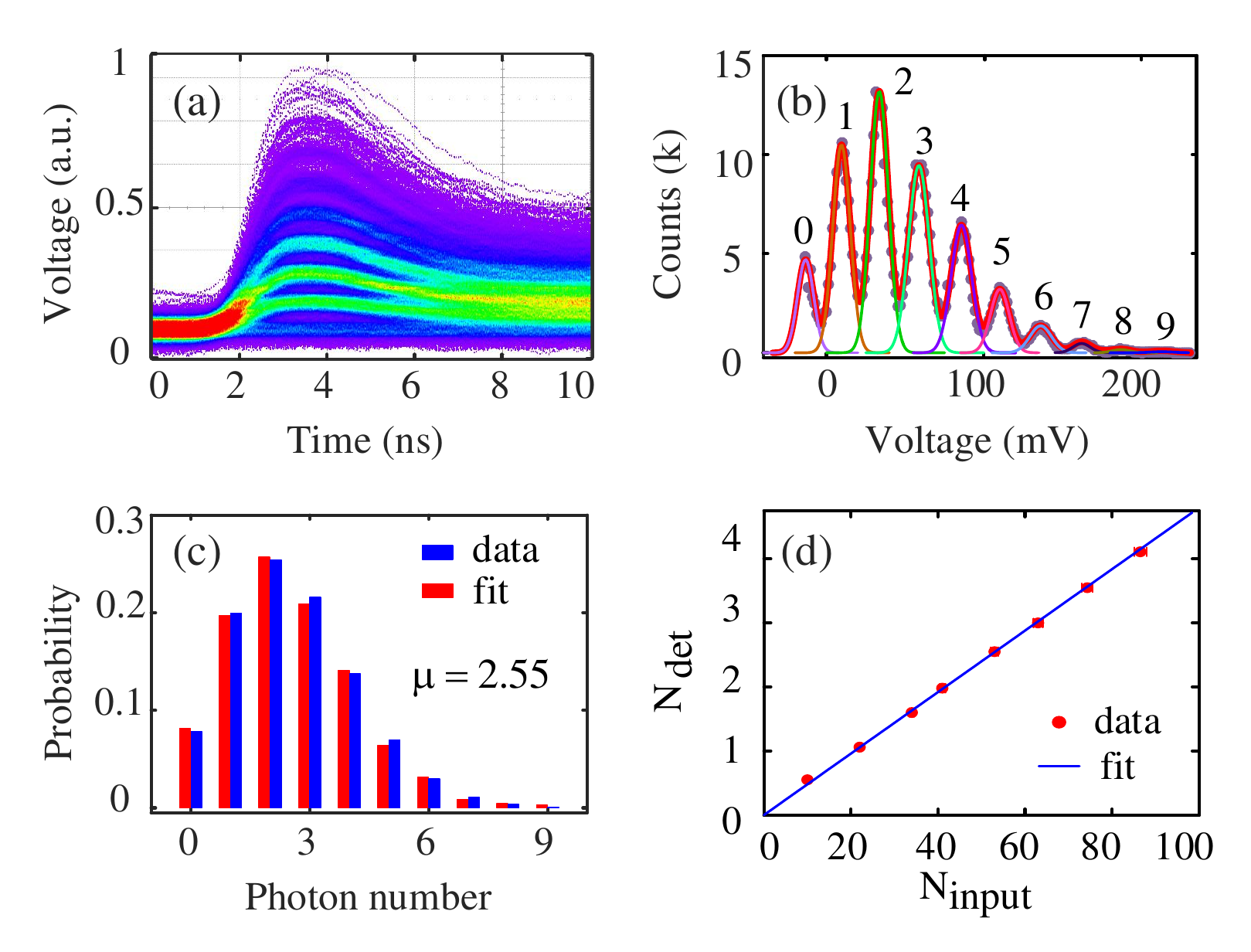}
\caption{Photon-number-resolving performance for the MIR frequency upconversion detector based on a Si-MPPC. (a) Superimposed waveforms of the MPPC signal recored by a digital oscilloscope. (b) Histogram of the photoresponse voltage amplitude for the upconverted SFG pulses. The solid lines represent the Gaussian fits while the numbers denote the corresponding photon-number states. (c) Photon-number distribution reconstructed from the results shown in (b). The obtained statistics is fitted by Poisson distribution, indicating an average photon number of 2.55 per pulse. (d) Detected photon number as a function of the incident photon number. The slope of the fitted linear line implies a detection efficiency of 4.8\%.}
\label{fig5}
\end{figure}

In order to demonstrate the PNR capability, the SPCM was replaced by a silicon-based multi-pixel photon counter (MPPC). The MPPC (Hamamatsu Photonics S13362-3050DG) was comprised of 60 $\times$ 60 pixels on an effective active area of 9 mm$^2$. The underlying mechanism for the PNR detection was based on the spatial multiplexing where incident photons would randomly hit different pixels. The breakdown voltage was optimized at 53 V for approaching a trade-off among quantum efficiency, dark noise, and photon-number resolution. The detection efficiency $\eta_\text{MPPC}$ was about 8\% at 771 nm with a dark noise about 15 kHz. The electronics for readout and processing were optimized for a high-repetition-rate operation at 14.6 MHz, which was much higher than previously reported values below several MHz \cite{Pomarico2010OE, Huang2011OL}. Figure \ref{fig5}(a) shows the typical waveforms recorded by a digital oscilloscope at the persistence display mode. The sampling rate was set at 20 GHz, which was sufficient to identify the photon clicks within a temporal duration about 4 ns. The exhibiting discrete amplitudes correspond to various detected photon numbers. Figure \ref{fig5}(b) gives the histogram of the voltage amplitude for the acquired 500,000 MPPC waveforms. Different photon-number states could be differentiated by the peaks in the histogram, leading a maximum resolved photon number up to 9. These peaks were then fitted by a series of Gaussian functions. The area under each peak after being normalized to the total one represents the photon-number probability. The reconstructed photon-number distribution is given in Fig. \ref{fig5}(c).

Generally, the measured probability $Q(n)$  for $n$ photons in the pulse is related to the incoming distribution $S(m)$ by the relation: $Q(n) = \sum_{m \geq n} P(n|m) \times S(m)$, where $P(n|m)$ is the conditional probability that $n$ photons are detected at the presence of $m$ incident photons. In our proof-of-principle experiment, the MIR signal was in a coherent state, thus following the Poissonian nature for the photon-number statistics. In this case, the detected state after losses was still a coherent state but with a smaller average photon number scaled by the overall detection efficiency. As shown in Fig. \ref{fig5}(c), the photon-number distribution was close to a poissonian $Q(n)=\mu^n \exp(- \mu) / n!$ with a detected average photon number $\mu$ = 2.55. The linear scaling between the input and detected photon numbers per pulse was manifested in Fig. \ref{fig5}(d), showing a large dynamic range for the incident MIR power. The slope of the linear fitting line indicates a total detection efficiency of 4.8\%, which includes the conversion efficiency, filtering transmission and detection efficiency of MPPC. The detection efficiency could be further increased by choosing a MPPC with a higher responsivity optimized at the SFG wavelength \cite{Huang2011OL}. Thanks to the low-noise upconversion system and high-speed repetition rate, the achieved noise probability was as low as $1.4 \times 10^{-3}$, which would impose negligible effect on the PNR performance. In our experiment, the background noise for the upconversion PNR detector was mainly limited by the dark noise for the detector itself. The intrinsic thermal noise could be minimized by using Peltier cooling \cite{Huang2011OL}.

\section{Summary}
To conclude, we have demonstrated MIR photon counting and resolving performance based on efficient nonlinear frequency upconversion. The employed pulsed pumping scheme leveraged the high peak power and ultrashort excitation window, which would help to increase the conversion efficiency and reduce the background noise. Thanks to the spectro-temporal engineering and tight passive synchronization of the involved light pulses, the quantum conversion efficiency was optimized up to 80\%. In combination of effective filtering stage and high-performance Si-SPCM, a record-high overall detection efficiency of 37\% was obtained with a corresponding NEP as low as 1.8$\times 10^{-17} \text{W/Hz}^{1/2}$. The presented upconversion detector was thus featured with ultra-high sensitivity at the room-temperature operation. Furthermore, the upconversion detector was extended with a Si-MPPC to demonstrate heretofore uninvestigated PNR capability at MIR. The maximum resolved photon number reached to 9, which enabled to implement a faithful reconstruction of Poissonian statistics for MIR coherent states. The achieved MIR photon recording and resolving capabilities would be promising to facilitate subsequent classical and quantum applications requiring single-photon sensitivity and large dynamic range, such as high-precision remote positioning \cite{Widarsson2020AO}, ultra-sensitive gas sensing \cite{Wolf2017OE}, and photon-number-thresholded LIDAR \cite{Cohen2019PRL, Bao2014AO}.

\section*{Funding} 
Science and Technology Innovation Program of Basic Science Foundation of Shanghai (18JC1412000), Program for Professor of Special Appointment (Eastern Scholar) at Shanghai Institutions of Higher Learning, National Key Research and Development Program (2018YFB0407100), National Natural Science Foundation of China (11621404, 11727812), Shanghai Municipal Science and Technology Major Project (2019SHZDZX01).

\section*{Disclosures} 
The authors declare no conflicts of interest.

\end{document}